\documentclass[aps,pra,amsmath,amssymb,reprint,superscriptaddress,floatfix]{revtex4-1}

\usepackage{xcolor}
\usepackage[]{graphicx}
\usepackage{xspace}
\usepackage{dsfont}
\usepackage{placeins}
\usepackage{hyperref}
\renewcommand{\vec}[1]{\ensuremath{\mathbf{#1}}} % for vectors
\newcommand{\ket}[1]{\vert #1 \rangle} % for Dirac bras
\newcommand{\ketbra}[2]{\left| #1 \middle> \! \middle< #2
\right|} % for operators
\newcommand{\abs}[1]{\vert #1 \vert} % for absolute value
\newcommand{\avg}[1]{\langle #1 \rangle} % for average
\newcommand{\unit}[1]{\ensuremath{\, \mathrm{#1}}}
\newcommand{\commutator}[2]{[ #1 , #2 ]} % for commutators
\newcommand{\br}[1]{\left( #1 \right)}
\newcommand{\brr}[1]{\left[ #1 \right]}
\newcommand{\brrr}[1]{\left\{ #1 \right\}}
\newcommand{\pd}[2]{\frac{\partial #1}{\partial #2}} % for partial derivatives
\DeclareMathOperator{\Tr}{Tr}
\newcommand{\Lio}{\mathcal{L}} % Liouvillian symbol

\renewcommand{\Re}{\mathrm{Re\,}}

\renewcommand{\Im}{\mathrm{Im\,}}

\newcommand*{\eg}{e.g.\@\xspace}
\newcommand*{\ie}{i.e.\@\xspace}

\renewcommand{\P}{\mathcal{P}}
\begin{document}

% Title of paper
\title{Laser and cavity cooling of a mechanical resonator with a\\
  Nitrogen-Vacancy center in diamond}

\author{Luigi Giannelli} \author{Ralf Betzholz}
\affiliation{Theoretische Physik, Universit\"at des Saarlandes, 66123
  Saarbr\"ucken, Germany} \author{Laura Kreiner}
\affiliation{Experimentalphysik, Universit\"at des Saarlandes, 66123
  Saarbr\"ucken, Germany} \author{Marc Bienert}
\affiliation{Theoretische Physik, Universit\"at des Saarlandes, 66123
  Saarbr\"ucken, Germany} \author{Giovanna Morigi}
\affiliation{Theoretische Physik, Universit\"at des Saarlandes, 66123
  Saarbr\"ucken, Germany}

% \email[]{Your e-mail address} \homepage[]{Your web page} \thanks{}
% \altaffiliation{}
\date{\today}

\begin{abstract}
  We theoretically analyse the cooling dynamics of a high-Q mode of a
  mechanical resonator, when the structure is also an optical cavity
  and is coupled with a NV center.  The NV center is driven by a laser
  and interacts with the cavity photon field and with the strain field
  of the mechanical oscillator, while radiation pressure couples
  mechanical resonator and cavity field. Starting from the full master
  equation we derive the rate equation for the mechanical resonator's
  motion, whose coefficients depend on the system parameters and on
  the noise sources. We then determine the cooling regime, the cooling
  rate, the asymptotic temperatures, and the spectrum of resonance
  fluorescence for experimentally relevant parameter regimes.  For
  these parameters, we consider an electronic transition, whose
  linewidth allows one to perform sideband cooling, and show that the
  addition of an optical cavity in general does not improve the
  cooling efficiency. We further show that pure dephasing of the NV
  center's electronic transitions can lead to an improvement of the
  cooling efficiency.
\end{abstract}

% \keywords{}

% \maketitle must follow title, authors, abstract, \pacs, and
% \keywords
\maketitle

\section{Introduction} %\section{\label{}}

Colour centers in diamond are widely studied because of their
exceptional properties as bright solid-state quantum emitters at room
temperature~\cite{Schroeder2016,Doherty2013}. Their dynamics is being
analysed in a wide variety of setups, which for instance can achieve
the strong coupling with high-finesse optical
resonators~\cite{Albrecht2013,Brouri2000,Kurtsiefer2000} and/or the
strain coupling with high-Q vibrating
structures~\cite{Brouri2000,Kurtsiefer2000,Teissier2014,Ovartchaiyapong2014,Kipfstuhl2014a,Li2015,Lee2016}
or standing mechanical
waves~\cite{MacQuarrie2013,MacQuarrie2015}. This experimental progress
makes NV centers promising candidates for realizing quantum hybrid
devices, namely, devices capable of interfacing photons, phonons, and
spin excitations in a controlled way, and can offer a wide range of
applications for quantum information
processing~\cite{Wrachtrup2006,Childress2013,Nemeto2014,Aspelmeyer2014}
and quantum
sensing~\cite{Wrachtrup2013,Hong2013,Schirhagl2014,Rondin2014,Aspelmeyer2014}.
It thus calls for identifying the perspectives for control of these
hybrid devices, which requires a systematic characterization of their
dynamics.

\begin{figure}[!htb]
  \includegraphics[width=\linewidth]{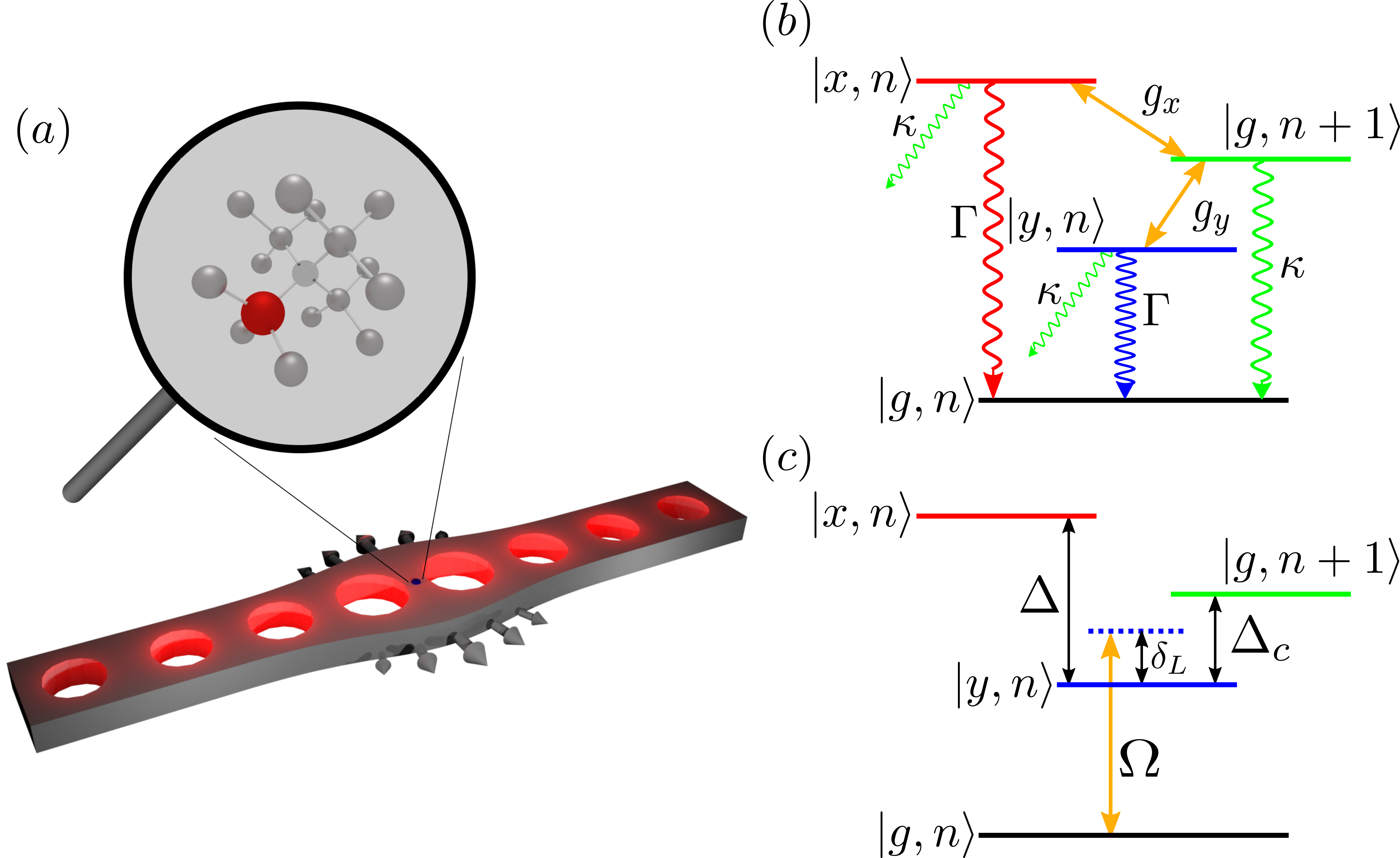}
  \caption{ \label{fig:1} (Color online) (a) A mechanical resonator,
    which is also a photonic crystal, interacts with a NV center in a
    diamond bulk via strain coupling.  (b) The NV-center internal
    level structure, including the photonic excitations: the ground
    state $\ket{g}\equiv\ket{{}^3A_{20}}$ couples to the excited
    states $\ket{x}\equiv\ket{E_x}$ and $\ket{y}\equiv\ket{E_y}$,
    which radiatively decay at rate $\Gamma$.  A mode of the
    high-finesse optical cavity decays at rate $\kappa$ and drives
    quasi-resonantly the transitions
    $\ket{g,n+1}\to\ket{x,n},\ket{y,n}$ with $n$ the intracavity
    photon number. Coefficients $g_x$ and $g_y$ denote the
    corresponding vacuum Rabi frequencies. (c) Sketch of the relevant
    frequencies $\delta_\text{L},\Delta_\text{c},\Delta$, as a
    function of which the cooling efficiency is characterised, in
    presence of a laser driving the transition $\ket{g}\to\ket{y}$
    with Rabi frequency $\Omega$.}
\end{figure}

In this work, we theoretically analyse laser cooling of a high-Q
vibrating mode, which is strain coupled to the electronic transitions
of an NV-center in diamond and optomechanically coupled to an optical
cavity. This situation can be realised, for instance, when NV center,
high-Q mechanical mode, and optical resonators are assembled in a
monolithic diamond structure, as illustrated in Fig.~\ref{fig:1} and
recently discussed in Refs.~\cite{Kipfstuhl2014a,Li2015}.  In this
setup the high-Q vibrating mode can be optomechanically cooled by the
coupling with the cavity and/or laser-cooled by the strain-coupling
with the NV-center transitions between the state
$\ket{g}\equiv\ket{{}^3A_{20}}$ and the levels
$\ket{E_x}\equiv\ket{x}$ and $\ket{E_y}\equiv\ket{y}$, sketched in
Fig.~\ref{fig:1}(b). The starting point of our study is the
theoretical model of Ref.~\cite{Kepesidis2013}, where the authors
investigated the effect of the NV multilevel structure on the dynamics
of a high-Q vibrational mode. We extend this model by including the
high-finesse mode of an optical cavity, which couples to the
electronic transitions of the NV center and to the mechanical
resonator by means of radiation pressure, and determine the laser
cooling dynamics.  We focus in particular on the regime where the
linewidth of the resonances induced by the coupling with the cavity is
of the same order as the one of the electronic transitions of the NV
center.  We further determine the effect of pure dephasing, which
tends to destroy the coherence of the NV-center excitations, on the
cooling dynamics. Surprisingly, we identify regimes where pure
dephasing can improve the cooling rate.

This article is organised as follows. In Sec.~\ref{sec:cav} we
  review some general concepts ruling the cooling dynamics in presence
  of the strong coupling with a resonator. In Sec.~\ref{sec:sys} the
theoretical model is introduced and in Sec.~\ref{sec:parameters} the
parameter regime is discussed with reference to existing experimental
realisation. In Sec.~\ref{sec:cooling} the rate equations for the
phonon dynamics are derived and in Sec.~\ref{sec:results} the cooling
rate, the asymptotic temperature, and the spectrum of resonance
fluorescence are determined and discussed in the presence and in the
absence of the coupling with the optical cavity mode. Moreover, the
cooling efficiency as a function of the dephasing rate is
analysed. The conclusions are drawn in Sec.~\ref{sec:conclusions}.

\section{General considerations}
\label{sec:cav}

Our study is motivated by an experimentally existing platform, like
the one sketched in Fig.~\ref{fig:1}. Our purpose is to investigate
whether the optomechanical coupling can help in achieving lower
temperatures than the ones that have been predicted by sideband
cooling using the strain-coupling with the NV center, see
Ref.~\cite{Kepesidis2013}. In fact, there can be an advantage by
resorting to the optical cavity if the final occupation of the
mechanical oscillator is lower than by just performing sideband
cooling with the NV center, and thus if (i) the cavity-assisted
cooling processes are sufficiently faster than the thermalization with
the external environment and yet (ii) the final occupation of the
oscillator is smaller than the one obtained by solely employing
sideband cooling, according to a protocol like the one described in
Ref.~\cite{Kepesidis2013}.  This analysis draws from several works
where it was studied how the interplay between the mechanical effects
due to the coupling of an electronic transition with a laser and with
a cavity can increase the cooling efficiency of a mechanical
oscillator
\cite{Cirac:1995,Vuletic:2003,Zippilli:2005,Zippilli:2007}. There it
was found that ground state cooling can be achieved as long as the
mechanical oscillator frequency, here denoted by $\nu$, is larger than
either the linewidth of the electronic transition, $\Gamma$, or of the
optical resonator, $\kappa$.  The minimal final mechanical oscillator
occupation one can achieve is then controlled by the ratio between the
linewidth of the narrower resonance, which we denote here by
$\Gamma_{\rm min} = \min{(\kappa,\Gamma)}$, and $\nu$. Accordingly,
the cooling rate $\tilde{\Gamma}$ is slower and scales with
$\Gamma_{\rm min}$.

These dynamics can be often illustrated by means of a set of rate
equations for the occupations $p_n$ of the oscillator's state with $n$
excitations ($n=0,1,2,\ldots$)~\cite{Stenholm:1986}:
\begin{equation}
  \label{Eq:rate}
  \dot p_n=-n(A_++A_-)p_n+(n+1)A_-p_{n+1}+nA_+p_{n-1}\,,
\end{equation}
with $\sum_np_n=1$ (see Sec.~\ref{sec:cooling} for details how this
equation is derived). Here, $nA_+$ and $nA_-$ are the rates with which
the oscillator in state $|n\rangle $ is heated and cooled,
respectively, by one phonon, and can have the form of lorentz
functions, whose linewidth is determined by the linewidth scattering
resonance.

Specific predictions for the relevant quantities, whose dynamics
Eq.~\eqref{Eq:rate} describes, can be directly extracted from the
equation for the mean phonon occupation number
$\avg{n} = \avg{a^\dag a}=\sum_nnp_n$~\cite{Stenholm:1986}:
\begin{equation}
  \dot{\avg{n}} = -(\tilde{\Gamma} + \gamma)(\avg{n} - n_f).
\end{equation}
Here $\gamma$ is the thermalization rate and $n_f$ the final phonon
occupation of the mechanical mode. Finally
\begin{gather}
  \label{eq:cooling_rate}
  \tilde{\Gamma} = A_- - A_+
\end{gather}
is the cooling rate when $A_->A_+$, whose maximum amplitude scales as
$\tilde \Gamma \sim (\omega_r/\nu)\Gamma_{\rm min}$ with $\omega_r$
the frequency scaling the mechanical effects due to the coupling with
light (when these are due to the phase or intensity gradient of the
light wave, $\omega_r$ is the recoil frequency; Here,
$\omega_r\sim(\Lambda/\hbar)^2/\nu$, with $\Lambda$ the strength of
the strain coupling). In this regime and for $\gamma=0$ radiation
cools the vibrations to the asymptotic occupation $N_0$, which is
given by
\begin{gather}
  \label{eq:mean_phonon}
  N_0 = \frac{A_+}{A_- -A_+} = \frac{A_+}{\tilde{\Gamma}}\,,
\end{gather}
and whose minimum scales with $N_0\sim (\Gamma_{\rm min}/\nu)^2$.

In a solid-state environment, where the heating rate due to the
coupling with the external reservoir is not negligible, slowing down
the cooling dynamics can be detrimental. This is visible when
considering the final occupation:
\begin{equation}
  \label{eq:n_f}
  n_f = \frac{\tilde{\Gamma}}{\tilde{\Gamma}+\gamma} N_0 +\frac{\gamma}{\tilde{\Gamma}+\gamma} N_{\rm th}\,,
\end{equation} 
where $N_{\rm th}$ is the mean phonon occupation at the temperature of
the external reservoir. Thus, maximizing the ratio
$\Gamma_{\rm min}/\gamma$ and minimizing the ratio
$\Gamma_{\rm min}/\nu$ is crucial and limits the parameter interval
where cavity-assisted cooling can improve the efficiency.

From these considerations one can generally identify the regime where
the coupling with a resonator can increase the sideband cooling
efficiency. In fact, a large cavity decay rate such that
$\kappa>\nu>\Gamma$ would increase the cooling rate
$\tilde{\Gamma}$. Yet it can also increase the asymptotic occupation
number of the mechanical mode $N_0$. On the other hand, a very good
cavity with $\kappa<\Gamma<\nu$ can allow one to achieve smaller
values of $N_0$, but at the price of decreasing $\tilde{\Gamma}$, so
that the final occupation number of the mechanical mode $n_f$ becomes
effectively larger.

The parameter regime to explore is quite large. However in general we
expect that, in the regime where laser sideband cooling is efficient,
the coupling to a resonator at linewidth $\kappa>\Gamma$ can be of
help only if it substantially increases the cooling rate keeping
$N_0<1$. The coupling to a resonator with $\kappa<\Gamma<\nu$ can help
in reaching ultralow temperatures, provided thermalization can be
neglected.  In this article we limit our analysis by taking the
optimal parameters for sideband cooling of an NV center and adding the
coupling with a cavity with linewidth $\kappa\sim \Gamma$, in order to
search for possible effects which cannot be foreseen drawing from
these simple considerations.  We refer the reader to
Sec.~\ref{sec:parameters} where the choice of the parameter regime is
discussed in relation to existing experimental implementations.  The
cooling rate, the asymptotic temperature, and the spectrum of
resonance fluorescence are then determined and discussed in
Sec.~\ref{sec:results} in the presence and in the absence of the
coupling with the optical cavity mode. The reader who is solely
interested in the resulting cooling efficiency can skip
Sec.~\ref{sec:cooling} and jump directly to Sec.~\ref{sec:results}.

\section{\label{sec:sys}The system}

In this Section we introduce the theoretical model which is at the
basis of our study. We describe the interaction of a high-Q mechanical
resonator mode of a phononic crystal cavity, with a quantum emitter,
specifically, a NV center in diamond, and a high-finesse optical
resonator mode of a photonic crystal cavity. The NV center is
strain-coupled with the mechanical resonator and the electronic dipole
transitions strongly couple with the photonic mode. The mechanical
resonator, in turn, is optomechanically coupled to the photonic
cavity.  The interactions in this system are expected to be strongly
enhanced by the co-localization in a single structure ensuring a
perfect spatial overlap between the different degrees of freedom,
which is achieved by assemblance in a monolithic diamond structure
sketched in Fig.~\ref{fig:1}(a). The system is intrinsically
dissipative due to radiative decay of the electronic excitations and
optical cavity losses.  Additionally, the mechanical resonator couples
to an external thermal reservoir. We assume that it is continuously
driven by a laser field, which directly couples to an electric dipole
transition of the defect.  In what follows we define the master
equation governing the dynamics of the density matrix $\rho$, which
describes the state of the composite system composed by the NV center,
and the photonic and phononic resonators.

\subsection{Basic equations}

The dynamics of the hybrid system's density operator $\rho$,
describing the state of the system composed by the internal degrees of
freedom of the NV-center, of the optical cavity mode and of the
mechanical oscillator, is governed by the master equation
$\partial_t\rho=\Lio \rho$, where superoperator $\Lio$ is defined as
($\hbar=1$):
\begin{equation}
  \label{eq:mastereq}
  \Lio \rho =-i \commutator{H}{\rho} + \mathcal{L}_{\rm dis}\rho\,,
\end{equation}
and which will be conveniently reported in the reference frame
rotating with the laser frequency $\omega_{\rm L}$. Below we provide
the detailed form of Hamiltonian $H$ and superoperator
$\mathcal{L}_{\rm dis}$.

\subsubsection{Unitary dynamics}

We first give the detailed form of the Hamiltonian $H$, which
generates the unitary part of the time evolution. For convenience, we
decompose it into the sum of Hermitian operators:
\begin{equation}
  \label{eq:hamiltonian}
  H = H_{\rm mec}+H_\text{I} + (a+a^\dag)V\,,
\end{equation}
where $a$ and $a^{\dagger}$ annihilate and create, respectively, a
mechanical vibration at frequency $\nu$, while $V$ acts on the cavity
and NV-center degrees of freedom and is specified later on. Operator
\begin{equation}
  H_{\rm mec} = \nu a^\dag a
\end{equation}
is the internal energy of the mechanical resonator, while Hamiltonian
$H_\text{I}$ describes the coupled dynamics of the NV center and the
optical cavity:
\begin{eqnarray}
  \label{H:I}
  H_\text{I}&=&\br{ \omega_y-\omega_{\rm L}}\ketbra{y}{y} +
                \br{\omega_x - \omega_{\rm L}}\ketbra{x}{x}+\br{\omega_{\rm c} - \omega_{\rm L}} c^\dag c\nonumber\\
            & &+ \left[\frac{\Omega}{2}\ketbra{y}{g} + \br{g_x\ketbra{x}{g} + g_y\ketbra{y}{g}}c+ {\rm H.c.}\right]\,.
\end{eqnarray}
Here, $\omega_x$ ($\omega_y$) is the frequency splitting in the
laboratory frame between the excited state $\ket{x}$ ($\ket{y}$) and
the ground state $\ket{g}$; operators $c$ and $c^{\dagger}$ annihilate
and create, respectively, a cavity photon at frequency
$\omega_{\rm c}$ (in the laboratory frame). The splitting between
  the $\ket{x}$ and $\ket{y}$ states is, for instance, due to a
  non-zero strain coupling, which is not related to the mechanical
  mode we consider. The frequencies appear shifted by
$\omega_{\rm L}$ since Hamiltonian $H_\text{I}$ is reported in the
reference frame rotating at the laser frequency. The second line of
Eq. \eqref{H:I} describes, from left to right, the external laser
driving the transition $\ket{g}\to\ket{y}$ with Rabi frequency
$\Omega$, while the optical mode drives the transitions
$\ket{g}\to\ket{x}$ and $\ket{g}\to\ket{y}$ with vacuum Rabi frequency
$g_x$ and $g_y$, respectively. We note that the laser polarization can
be chosen to selectively drive one electronic transition, as we do in
our model, while in general the cavity mode's polarization has a
finite projection to the dipole moment of both transitions, since this
depends on the preparation of the sample. Therefore, we generally
assume $g_x,g_y\neq 0$ unless otherwise stated. The relevant NV center
and cavity states are reported in Fig.~\ref{fig:1}(b)-(c) with the
relative detunings with respect to the laser frequencies. These are
defined as:
\begin{equation}
  \label{eq:defdeltas}
  \begin{aligned}
    &\delta_{\rm L} = \omega_{\rm L} - \omega_{\rm y}, \\
    &\Delta = \omega_{\rm x} - \omega_{\rm y}, \\
    &\Delta_{\rm c} = \omega_{\rm c}- \omega_{\rm y}\,.
  \end{aligned}
\end{equation}

Finally, operator $V$ is the sum of the strain and of the
optomechanical coupling of the mechanical resonator with NV center and
optical cavity, respectively. We decompose it hence into the sum
$V=V_{\rm strain}+V_{\rm om}$, where $V_{\rm strain}$ acts on the NV
degrees of freedom and reads~\cite{Fu2009}
\begin{equation}
  V_{\rm strain} = \sum_{j=I,X,Z}\Lambda_j A_j \,,
\end{equation}
where $\Lambda_j$ are the strain coupling constants and the operators
$A_j$ are defined as:
\begin{equation}
  \begin{aligned}
    \label{eq:As}
    A_I &= \ketbra{x}{x} + \ketbra{y}{y}\,, \\
    A_X &= \ketbra{x}{y} + \ketbra{y}{x}\,, \\
    A_Z &= \ketbra{x}{x} - \ketbra{y}{y}\,.
  \end{aligned}
\end{equation}
The optomechanical coupling reads $V_{\rm om}=- \chi c^\dag c$ with
$\chi$ the optomechanical coupling constant \cite{Law1994,Law1995}.

\subsubsection{Dissipation}

The irreversible processes we consider in our theoretical description
are: (i) the radiative decay of the NV excitations and pure dephasing
of the electronic coherences, (ii) cavity losses, and (iv) the
mechanical damping rate due to the coupling of the mechanical
resonator with an external thermal reservoir. We model each of these
phenomena by a Born-Markov process described by the corresponding
superoperator, such that superoperator $\mathcal{L}_{\rm dis}$ in
Eq. \eqref{eq:mastereq} can be cast in the form
\begin{align}
  \mathcal{L}_{\rm dis}=\Lio_{\Gamma} + \Lio_\kappa +\Lio_\gamma\,.
\end{align}
The individual terms read
\begin{align}
  \label{eq:NVliouvillian}
  \mathcal{L}_\Gamma &= \frac{\Gamma}{2} \sum_{\xi = x, y} \mathcal{D}[\ketbra{g}{\xi}]
                       + \frac{\Gamma_\phi}{2} \sum_{\xi = x, y} \mathcal{D}[\ketbra{\xi}{\xi}]\,,\\
  \mathcal{L}_\kappa &= \frac{\kappa}{2}\mathcal{D}[c]
                       \,,\\  \label{eq:mechanicalliouvillian}
  \mathcal{L}_\gamma &= \frac{\gamma}{2}\left(N_{\rm th}+1 \right)\mathcal{D}[a] + \frac{\gamma}{2}N_{\rm th}\mathcal{D}[a^\dag]\,,
\end{align}
where we used the definition
\begin{equation}
  \mathcal{D}[o]\rho =  2o\rho o^\dag -o^\dag o\rho -\rho o^\dag o\,,
\end{equation}
with $o=\ketbra{g}{\xi},\ketbra{\xi}{\xi},c,a,a^\dag$. The
coefficients are the radiative decay rate $\Gamma$ of the NV-center
excited states, the dephasing rate of the electronic coherences
$\Gamma_\phi$, cavity losses at rate $\kappa$, and the damping rate of
the mechanical oscillator $\gamma$. Finally,
$N_{\rm th} = \left(\exp(\nu / k_{\rm B} T) -1 \right)^{-1}$ is the
equilibrium phonon occupation number of the bath to which the
oscillator couples, with $T$ the bath's temperature.

\subsection{Spectrum of resonance fluorescence}

In what follows we will use the master equation,
Eq. \eqref{eq:mastereq}, in order to analyze the cooling efficiency of
the mechanical resonator and the spectrum of the light emitted by the
NV center at the steady state of the cooling dynamics. In order to
better characterize the parameter regime where cooling is efficient we
choose an analytical approach, which is based on a perturbative
expansion of the Lioville operator and allows us to determine the
cooling regime, the corresponding rate and the asymptotic
temperature. This approach is reported in the following Section.

Moreover, in the regimes of interest we determine the spectrum of the
scattered light, for the purpose of identifying the relevant features
in the photons which are emitted by the NV center outside of the
resonator. The spectrum of resonance fluorescence is, apart from a
constant proportionality factor, the Fourier transform of the
auto-correlation function of the electric field~\cite{Glauber2007}:
\begin{equation}
  \label{eq:spectrum}
  \mathcal{S}(\omega)\propto{\rm Re}\int_0^{\infty}{\rm d}\tau\,{\rm
    e}^{-i\omega\tau} \langle E^{(-)}(\tau)E^{(+)}(0)\rangle_{\rm st} 
\end{equation}
where $E^{(-)}(t)$ and $E^{(+)}(t)$ are the negative and positive
frequency component of the electric field at time $t$ and
$\langle\cdot\rangle_{\rm st}\equiv {\rm Tr}\{\cdot\rho_{\rm st}\}$
denotes the trace taken over the steady state density matrix
$\rho_{\rm st}$ which solves $\Lio\rho_{\rm st}=0$. The intensity of
the scattered field (away from the forward direction) is proportional
to the source field, hence in the far-field the electric field is
proportional to the sum of the operators
$\vec{d}_j\ketbra{g}{j}+{\rm H.c.}$, for $j=x,y$ where $\vec{d}_x$ and
$\vec{d}_y$ are the dipole moments of the transitions
$\ket{g}\to\ket{x}$ and $\ket{g}\to\ket{y}$, respectively (notice that
$\abs{\vec{d}_x} = \abs{\vec{d}_y}$). Since the dipole moments are
mutually orthogonal, the spectrum integrated over the full solid angle
$4\pi$ is the incoherent sum of the two components coming from the
$\ketbra{g}{x}$ and $\ketbra{g}{y}$ operators, \ie the interference
term integrates to zero. With the help of the quantum regression
theorem~\cite{carmichael1} one can cast the spectrum into the form
\begin{equation}
  \label{eq:spectyum1}
  \mathcal{S}(\omega) \propto \sum_{j=x,y}\Re{\Tr{\brrr{\ketbra{j}{g}
        \brr{i(\omega-\omega_{\rm L}) -\Lio}^{-1} \ketbra{g}{j} \rho_{\rm st}}}}.
\end{equation}
In this work we numerically determine the spectrum for the parameter
regimes of interest.

\section{Parameter regime}
\label{sec:parameters}

In order to justify the experimental relevance of the cooling dynamics
we discuss in the rest of this article, we now relate the theoretical
model to existing experimental realisations and identify the parameter
regime which we will consider in our analysis.

{\it Optical resonator}. A structure like the one discussed here can
be found for instance in a so-called phoxonic crystal (PxC), which
co-localizes confined optical and mechanical resonator
modes~\cite{Kipfstuhl2014a}. Photonic crystals are formed by a
periodic modulation of the refractive index (in this case air holes in
diamond), resulting in the formation of optical bands similar to
electronic band structures in solids. A local defect like \eg a
variation of the hole diameters along the PxC structure perturbs the
perfect periodicity and gives rise to an optical cavity mode. So far,
fabrication imperfections limit experimental quality factors to $10^4$
at visible wavelengths suitable for the interaction with colour
centers in diamond and up to $10^5$ in the telecom band around
$1550\unit{nm}$~\cite{Hausmann2013a,Burek2014,Li2015,Burek2015}. Nevertheless,
simulations of one-dimensional photonic crystal cavities designed for
visible light predict quality factors up to $10^7$ and mode volumes
around $1$ cubic wavelength with cavity loss rate in the range
$\kappa\sim\!  10\unit{MHz}-1\unit{GHz}$~\cite{Kipfstuhl2014a}.

{\it Mechanical resonator}: In a PxC a periodic variation of the
elastic modulus creates a mechanical band structure and a suitable
variation of the regular pattern allows for a localized mode of the
mechanical resonator.  Recent experiments with structures at
mechanical frequencies of $6\unit{GHz}$ with optical properties
suitable for telecom wavelengths show mechanical quality factors of
$10^3$~\cite{Burek2015}. Numerical modeling shows that modes with
frequencies in the range $10-20\unit{GHz}$ with quality factors
reaching $10^7$ can be achieved with an effective mass of
$10^{-16}\unit{kg}$ for structure dimensions matching visible
wavelengths with the confined optical mode~\cite{Kipfstuhl2014a}. The
parameters we choose are consistent with assuming mechanical
frequencies of the order of $1-10\unit{GHz}$ and a quality factor of
the order of $10^6-10^7$, giving a damping rate $\gamma$ of few
$\unit{kHz}$. The strain coupling constants are taken to be of the
order of
$1-10\unit{MHz}$~\cite{Kepesidis2013,Tamarat2006,Doherty2011}. The
optomechanical coupling constant $\chi$ is taken to be of the order of
few $\unit{MHz}$~\cite{Kipfstuhl2014a}.

{\it NV center}. Figure~\ref{fig:1}(b)-(c) reports the relevant level
structure of the NV center in diamond. In absence of strain coupling,
the $m_s=0$ ground state $\ket{{}^3A_{20}}$ can be selectively coupled
to the excited states $\ket{E_{x,y}}$, which have zero spin angular
momentum. While the ground state is much less sensitive to lattice
distortion, these excited states are highly susceptible to external
perturbations~\cite{Maze2011,Doherty2011,Doherty2013}. Axial strain
(parallel to the NV center axis, equivalent
$\left\langle 111 \right\rangle$ crystal direction) leads to an
additional splitting between ground and excited states as well as
between the $m_s=0$ and $m_s=\pm1$ levels in the ground state.  Radial
strain (perpendicular to NV axis) mixes the excited state levels $E_x$
and $E_y$ and leads to a splitting of the new states $E_x^*$ and
$E_y^*$ ($m_s=+1^*$ and $m_s=-1^*$). The effect of strain coupling on
the excited states is several orders of magnitude larger than on the
ground state and hence dominates the strain-induced modification of
the NV's optical properties. Therefore, we restrict our model to the
interaction between the mechanical resonator mode and the transition
coupling the ground state $\ket{g}\equiv\ket{{}^3A_{20}}$ to the
excited states $\ket{x}\equiv\ket{E_x}$ and
$\ket{y}\equiv\ket{E_y}$. For the excited states we take the radiative
decay rate
$\Gamma\!\sim\!100\unit{MHz}$~\cite{Sipahigil2012,Bernien2012}. The
interaction between the NV transitions and the $71\unit{meV}$ lattice
phonon modes~\cite{Gali2011} changes the energy of the $\ket{x}$ and
$\ket{y}$ states and can thus give rise to a dephasing mechanism of
the electronic coherence~\cite{Albrecht2013,Betzholz2014}. In our
model we neglect the mixing between the states and consider only pure
dephasing with rates of the order of $\Gamma_\phi \sim 100\unit{MHz}$,
which can be achieved in bulk diamond at temperatures lower than
$10\unit{K}$~\cite{Abtew2011,Fu2009}.  We restrict the frequency of
the mechanical resonator mode to $\nu = 2\pi\times 1\unit{GHz}$ in
order to avoid coupling to NV excited states other that $E_x$ and
$E_y$. As the optical cavity mode should still be near resonant on the
optical transition of the NV at $637\unit{nm}$ this doesn't correspond
to a real structure design for the full threefold
hybrid-system. However, we still model this artificial parameter set
in order to obtain qualitative results on the nature of the
interaction.

{\it Cooling regime}: The analysis of the cooling efficiency is
performed by determining the cooling rate $\tilde{\Gamma}$ and the
ideal asymptotic occupation number of the mechanical mode $N_0$ as a
function of the tunable parameters, which we take here to be the
frequency splitting of the electronic excited states and the laser
frequency, corresponding to changing $\delta_{\rm L}$,
$\Delta_{\rm c}$, and $\Delta$.  The analysis is performed by
searching for the parameter regime where the asymptotic occupation
number $N_0<1$ and the cooling rate $\tilde{\Gamma}$ is maximized, in
order to realise regimes where the radiative cooling can overcome
thermalization by the external reservoir. This constrains the range of
parameters. A necessary condition for performing ground state cooling
is the presence of a resonance whose linewidth $L$ is smaller than the
trap frequency \cite{Eschner:2003}, which poses an upper bound to
$L$. Moreover, the cooling rate shall exceed the thermalization
rate. Since the cooling rate is proportional to the effective
linewidth of the cooling transition, this condition sets a lower bound
to $L$. If one performs optomechanical cooling by driving the optical
resonator, then $L=\kappa$. In absence of the resonator, the
mechanical oscillator can be cooled by driving the NV center
transitions with a laser and $L=\Gamma$.  When the NV center
transitions also couple with the optical cavity, then $L$ is a linear
interpolation of the cavity linewidth $\kappa$ and of the NV
transition linewidth $\Gamma$, and varies between $\Gamma$ and
$\kappa$ \cite{Zippilli:2005} (smaller linewidths could be achieved by
coupling to other stable electronic transitions, which in our system
are not considered \cite{EIT:2000,CEIT:2012}).

In order to get a relatively small phonon occupation of the bath
$N_{\rm th}$ we take a large mechanical frequency,
$\nu\sim 2\pi\times 1\unit{GHz}$, and thus for our parameter choice
$\Gamma<\nu$.  We then fix the cavity linewidth $\kappa\simeq\Gamma$.

\section{Effective dynamics of the mechanical resonator}
\label{sec:cooling}

For the parameter regime we consider all characteristic frequencies
characterizing the coupling of the mechanical resonator with NV center
and optical cavity are much smaller than the mechanical resonator
eigenfrequency
$\Lambda_{I},\Lambda_{X},\Lambda_{Z},\chi\bar n_{\rm c}\ll \nu$
($\bar n_{\rm c}$ being the mean intracavity photon occupation
number). This justifies a perturbative treatment, which allows us to
eliminate the degrees of freedom of NV and optical cavity from the
dynamics of the mechanical oscillator in second-order perturbation
theory. By means of this procedure we derive an effective master
equation for the mechanical resonator only, which allows us to
determine the parameter regime where the vibrations are cooled, the
corresponding cooling rate and the asymptotic vibrational state.

\subsection{\label{sec:pert}Perturbative expansion}

We derive a closed master equation for the mechanical oscillator
starting from Eq.~\eqref{eq:mastereq} and assuming that the coupling
frequencies, which scale the operator $a+a^\dag$, are much smaller
than $\nu$.  This can be summarized by the inequality $\alpha\ll\nu$,
with $\alpha=\Lambda_{I},\Lambda_{X},\Lambda_{Z},\chi\bar n_{\rm c}$
and $\bar n_{\rm c}$ the mean intracavity photon occupation number. We
then perform perturbation theory in second order in the small
parameter $\alpha/\nu$. We further assume that the incoherent dynamics
of the oscillator due to the coupling with the environment is
sufficiently slow that the occurrence of these processes during a
scattering process can be discarded. This requires that
$\gamma N_{\rm th}\ll \alpha$, which for the parameters considered in
this work is valid also at room temperature, so that we treat it in
first order.

According to these considerations we split the Liouville operator as
$$\Lio=\Lio_0+\mathcal{V}+\Lio_\gamma\,,$$ 
with $\Lio_0= \Lio_{\rm E}+ \Lio_{\rm I}$, where $\Lio_{\rm E}$ and
$\Lio_{\rm I}$ are the Liouville operators that generate the uncoupled
mechanical oscillator and internal (NV center + optical cavity)
dynamics, respectively, while $\mathcal{V}$ describes the coupling
between mechanical and internal degrees of freedom. In detail,
\begin{align}
  \label{eq:LiouvillianDecompose}
  \Lio_{\rm E}\rho&=-i [H_{\rm mec},\rho],\\
  \label{eq:LiouvillianDecompose2}
  \Lio_{\rm I}\rho &=-i [H_{\rm I},\rho] + \Lio_{\Gamma}\rho + \Lio_\kappa \rho,\\
  \label{eq:LiouvillianDecompose:L1}
  \mathcal{V}\rho &= -i [V(a+a^\dagger), \rho]\,.
\end{align}
We formally eliminate the coupling between mechanical resonator and
internal degrees of freedom as done for instance in
Refs.~\cite{Javanainen1984,Cirac1992,Morigi2003a,Zippilli:2005}. We
first introduce the superoperators $\mathcal{P}_k$ such that
\begin{align}
  \mathcal{P}_k\rho=\sigma_{\rm st}\sum_{n=0}^\infty|n\rangle\!\langle
  n+k|\langle n|\mu|n+k\rangle\,,
\end{align}
with $\mu=\Tr_{\rm I}\{\rho(t)\}$ the reduced density matrix,
$\Tr_{\rm I}\{\cdot\}$ being the trace over the internal degrees of
freedom, $\ket{n}$ the eigenstates of the mechanical oscillator,
$k=0,\pm1,\pm2,...\,$ ($k\ge -n$) and $\sigma_{\rm st}$ the steady
state for the internal degrees of freedom:
$\Lio_{\rm I}\sigma_{\rm st}=0$. Applying $\P_k$ to the master
equation~\eqref{eq:mastereq}, with the definitions of the
superoperators~\eqref{eq:LiouvillianDecompose}-\eqref{eq:LiouvillianDecompose:L1},
in a second-order perturbative expansion in parameter $\alpha/\nu$ and
first order in $\gamma (N_{\rm th}+1)$, leads to the equation
\begin{align}
  \label{eq:effeqwithprojector}
  \pd{}{t}\mathcal{P}_k \rho&=\Big\{ik\nu + \P_k\mathcal{V}\big(ik\nu
                              - \Lio_0\big)^{-1}\mathcal{Q}_k\mathcal{V}\P_k\Big\}\P_k\rho+\Lio_{\gamma}\mathcal{P}_k \rho\,,
\end{align}
with $\mathcal{Q}_k=1-\P_k$ and $1$ is here the superoperator whose
action is the identity on both sides of the density matrix. The master
equation for the reduced density matrix $\mu$ is obtained after
tracing out the internal degrees of freedom in
Eq.~\eqref{eq:effeqwithprojector} and reads
\begin{equation}
  \label{eq:effectivedynmecexplit}
  \begin{aligned}
    \dot\mu = -i\bar{\nu}\commutator{a^\dag a}{\mu}+
    \frac{A_-}{2}\mathcal{D}[a]\mu +
    \frac{A_+}{2}\mathcal{D}[a^\dag]\mu+\Lio_{\gamma}\mu\,.
  \end{aligned}
\end{equation}

The rates $A_\pm$ are defined as
\begin{gather}
  \label{eq:A+A-nubar}
  A_\pm = 2\,\Re{s(\mp \nu)}, \\
  \bar{\nu} = \nu+\Im{s(\nu)} + \Im{s(-\nu)}\,,
\end{gather}
with
\begin{equation}
  \label{eq:sofv}
  s(v) = \int_0^\infty dt\, e^{ivt}\avg{V\exp{\br{\Lio_{\rm I} t}}V}_{\rm st}\,,
\end{equation}
which is the Fourier component at frequency $\nu$ of the
autocorrelation function of operator $V$, where the average
$\avg{\cdot}_{\rm st}$ is taken in the steady state $\sigma_{\rm st}$.

% \subsection{\label{sec:effcool}Rate equations}

The diagonal elements of Eq. \eqref{eq:effectivedynmecexplit} give a
set of rate equations for the occupation $p_n=\langle n|\mu|n\rangle$
of the phonon state $\ket{n}$, which are reported in
Eq. \eqref{Eq:rate}.

\section{Results}
\label{sec:results}

In this section we characterize the parameter regimes in which the
mechanical resonator is cooled by photon scattering process in the
setup of Fig.~\ref{fig:1}(a). We focus on the range of parameters
discussed in Sec. \ref{sec:parameters}.  We consider laser cooling of
the mechanical resonator by strain coupling with the NV center and
analyse how the cooling dynamics is affected by the presence of the
optical resonator and of dephasing. The results we report are compared
to the predictions in absence of the optical resonator and for
vanishing dephasing.  This latter case has been extensively discussed
in Ref. \cite{Kepesidis2013} and we refer the interested reader to it
for a detailed discussion of the predicted dynamics in that specific
limit.

\subsection{Cavity-assisted cooling}

We now analyse how laser cooling dynamics of the mechanical resonator
by strain coupling with the NV center is affected by the presence of
the optical resonator. In order to better understand the role of the
resonator, we first discard thermal effects and dephasing (setting
$\gamma=\Gamma_\phi=0$).

\begin{figure}[!htb]
  \includegraphics[]{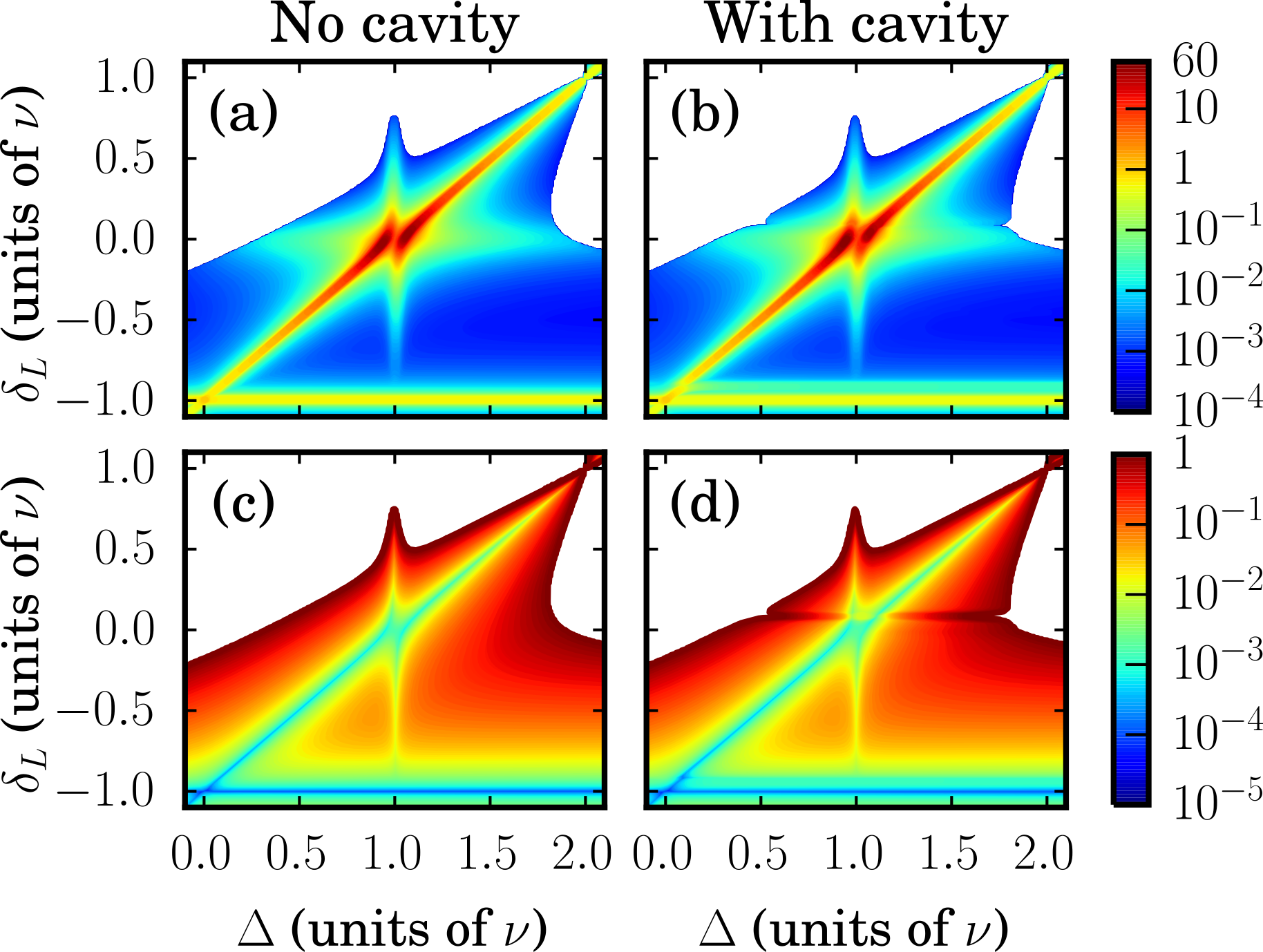}
  \caption{ \label{fig:2} (Color online) Predictions on the cooling
    efficiency extracted from the rate equation, Eq. \eqref{Eq:rate},
    for laser cooling of the mechanical resonator by driving the NV
    center with a laser (left panel) and by additionally coupling the
    dipole transitions to an optical cavity (right panel). (a) and (b)
    show the cooling rate $\tilde{\Gamma}$,
    Eq.~\eqref{eq:cooling_rate} in units of $\Lambda^2/\nu$, (c) and
    (d) the asymptotic occupation $N_0$ of the vibrational mode,
    according to Eq. \eqref{eq:mean_phonon}, as a function of the
    excited level splitting $\Delta$ and the laser detuning
    $\delta_{\rm L}$ (in units of $\nu$). The white region are heating
    regions ($\tilde{\Gamma}<0$) or where $N_0>1$. The parameters for
    the left panel are $\Omega=0.1\nu$, $\Gamma=1.6\times 10^{-2}\nu$,
    $\Gamma_\phi=0$,
    $\Lambda_I=0,\ \Lambda_X=\Lambda_Z=\chi=\Lambda=0.1\Gamma$ and
    $g_x=g_y=0$. In the right panel we take the same parameters except
    for $g_x=g_y=\kappa=\Gamma$. The cavity frequency is fixed to the
    value $\Delta_{\rm c}=8.5\times 10^{-2}\nu$ (see text).}
\end{figure}

\begin{figure}[!htb]
  \includegraphics[]{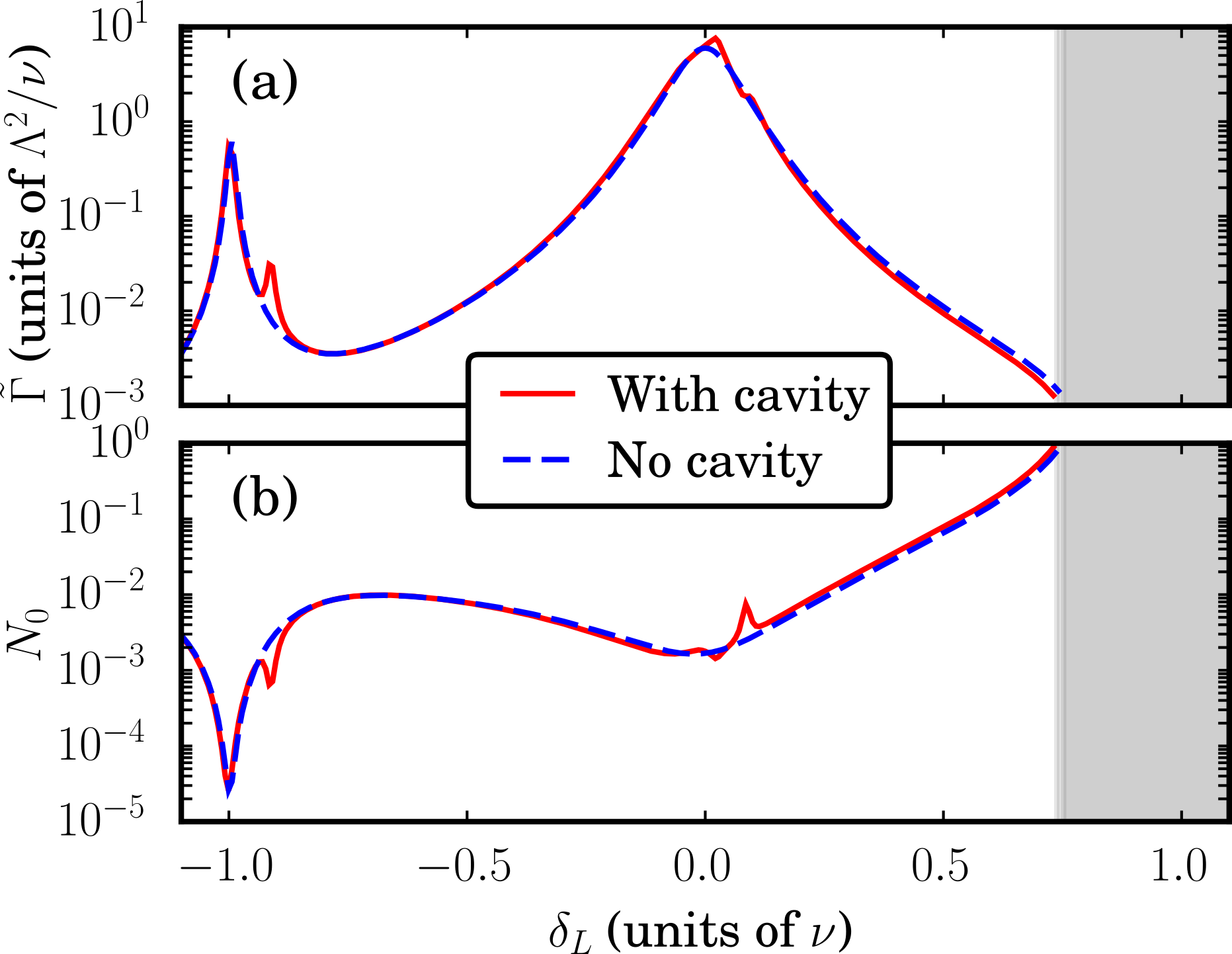}
  \caption{ \label{fig:3} (Color online) (a) Cooling rate
    $\tilde{\Gamma}$ and (b) asymptotic occupation $N_0$ of the
    vibrational mode as a function of $\delta_{\rm L}$ for the same
    parameters as in Fig.~\ref{fig:2} and $\Delta=\nu$. The dashed
    (solid) line corresponds to the predictions in absence (presence)
    of the coupling to the cavity. The shaded region indicates the
    regime where the resonator is heated by the radiative processes
    ($\tilde{\Gamma}<0$) or where $N_0>1$.  }
\end{figure}
 
Figure~\ref{fig:2} displays the cooling rate $\tilde{\Gamma}$ and the
mean vibrational number at the asymptotics $N_0$ as a function of
$\delta_{\rm L}$ and $\Delta$ in absence (left panels) and in presence
of the optical cavity (right panel). Both plots show that the cooling
rate is maximum, and the final occupation minimum, along the lines
$\delta_{\rm L}=-\nu$ and $\delta_{\rm L}=\Delta-\nu$. In the first
case cooling is achieved by setting the laser frequency to the value
$\omega_{\rm L}=\omega_y-\nu$, hence resonantly driving the transition
$|g,n\rangle\to |y,n-1\rangle$ (red sideband). In the second case the
laser frequency is $\omega_{\rm L}=\omega_x-\nu$, so that the
transition $|g,n\rangle\to |x,n-1\rangle$ is resonantly driven by an
effective process, which combines the laser and the strain
coupling. For most values of the detuning $\Delta$ the excitation of
the intermediate state $|y\rangle$ is virtual, except for
$\Delta=\omega_x-\omega_y=\nu$. This latter case corresponds to the
vertical line visible in both figures, where cooling results to be
efficient. These properties have been identified and discussed in
Ref.~\cite{Kepesidis2013} and do not depend on the coupling with the
resonator. The curves in Fig.~\ref{fig:3} show the cooling rate and
the minimum phonon occupation as a function of $\delta_{\rm L}$ after
fixing the detuning $\Delta=\nu$. Some (relatively small) differences
are visible close to the values $\delta_{\rm L}=0$ and
$\delta_{\rm L}=-\nu$, which are due to the level splitting induced by
the strong coupling with the resonator: for this choice of $\Delta_c$,
in fact, the cavity drives almost resonantly the transition
$|g\rangle\to |y\rangle$.

\begin{figure}[!htb]
  \includegraphics[]{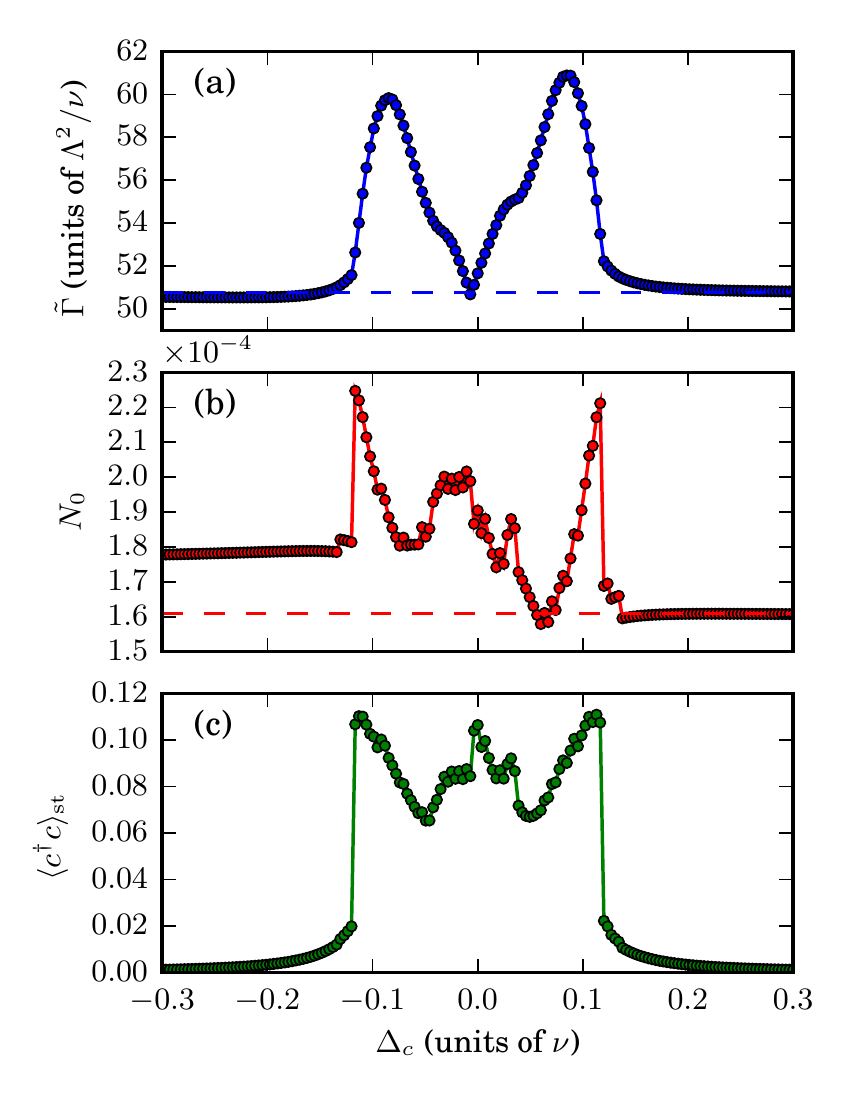}
  \caption{ \label{fig:4} (Color online) (a) Maximum cooling rate
    $\tilde{\Gamma}_{\rm max}$ in presence of the resonator as a
    function of $\Delta_{\rm c}$. The value $\tilde{\Gamma}_{\rm max}$
    has been calculated by varying $\delta_{\rm L}$ and $\Delta$ and
    keeping $\Delta_{\rm c}$ fixed.  Subplot (b) displays the
    corresponding value of $N_0$ and (c) the mean intracavity photon
    number. The parameters are: $\Omega=0.1\nu$, $\Gamma_\phi=0$,
    $\Gamma=\kappa=g_x=g_y=1.6\times10^{-2}\nu$,
    $\Lambda_I=0,\ \Lambda_X=\Lambda_Z=\chi=\Lambda=0.1\Gamma$. The
    dashed lines in (a) and (b) indicate the maximum cooling rate and
    corresponding value of $N_0$ in absence of the optical
    resonator. In the latter case $\tilde{\Gamma}_{\rm max}$ is
    maximum for $\Delta\approx0.93\nu$ and
    $\delta_{\rm L}\approx-3.5\times 10^{-2}\nu$.}
\end{figure}

We have tested that the value of the detuning $\Delta_{\rm c}$, and
thus of the cavity frequency, in Figs. \ref{fig:2} and \ref{fig:3},
leads to the best results by comparing cooling rate and final
temperature for different values of $\Delta_{\rm c}$. The results of
this analysis are summarized in Fig.~\ref{fig:4}, which displays (a)
the maximum cooling rate (maximized by varying $\Delta$ and
$\delta_{\rm L}$ by keeping $\Delta_c$ fixed).  The mean phonon
occupation in (b) and the mean intracavity photon number in (c) are
reported for the corresponding values of $\Delta$ and
$\delta_{\rm L}$, at which $\tilde{\Gamma}$ is maximum. These plots
show that maximal cooling rates are found for $\Delta_c\simeq 0$. We
verified that the curves do not differ substantially if instead we
search for $\Delta_{\rm c}$ by minimizing the mean phonon
number. Therefore, the contour plots in Fig. \ref{fig:2}(b) and (d)
show the optimal cooling rate and temperature in presence of the
resonator.  On the basis of the comparison with the plots on the left
panels, we can thus conclude that the coupling with the cavity does
not substantially improve the cooling efficiency for the chosen
parameter regime.

\begin{figure*}[!htb]
  \includegraphics[]{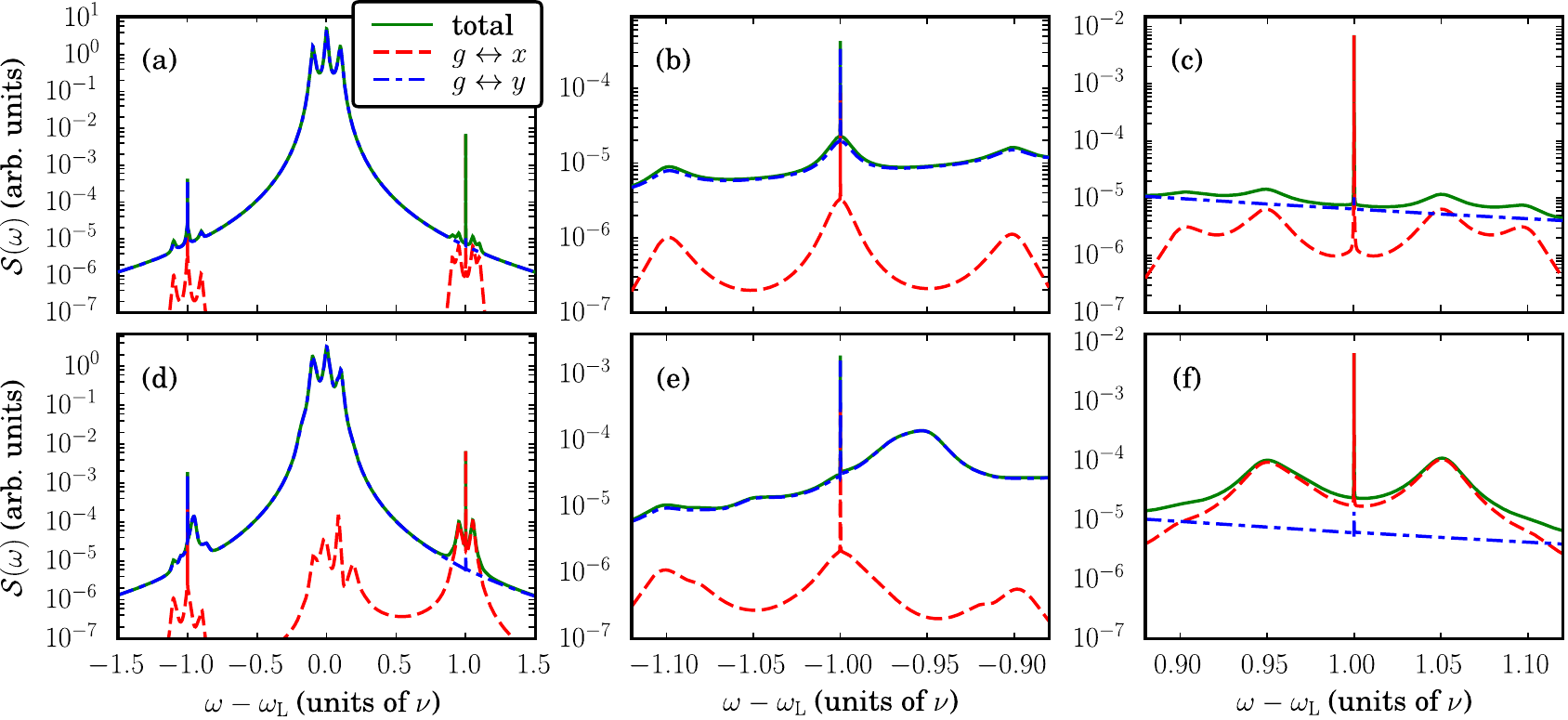}
  \caption{ \label{fig:5} (Color online) Spectrum of light emitted by
    the NV center at the asymptotics of the laser cooling
    dynamics. The upper panels correspond to the parameter regime of
    Fig.~\ref{fig:2}(a)(c) (no cavity), the lower panels to the
    parameter regime of Fig.~\ref{fig:2}(b)(d) (cavity assisted
    cooling). The dashed (dashed-dotted) line correspond to the
    emission from the transition $g \leftrightarrow x$
    ($g \leftrightarrow y$), the solid line correspond to the sum of
    these two contributions. Here, we took $\Delta=\nu$ and
    $\delta_{\rm L}=0$.  Panels (b), (c), (e) and (f) show the details
    of the sidebands.}
\end{figure*}

We now analyse how the spectrum of resonance fluorescence is modified
by the coupling with the resonator. We focus on the light emitted once
the system has reached the stationary state. Figure~\ref{fig:5}
displays the resonance fluorescence spectrum in absence and in
presence of the optical cavity for the parameters of Fig.~\ref{fig:3}
with $\delta_{\rm L}=0$. To better understand how the cavity modifies
the dynamics, we first discuss the spectrum in absence of the
cavity. In this case we observe the three broad resonances around
$\omega=\omega_{\rm L}$.  These are due to inelastic processes in
which the motion is not involved and can be interpreted as a
Mollow-type triplet~\cite{Mollow1969}.  We further observe the narrow
resonances at $\omega = \omega_{\rm L} \mp \nu$, which are the red and
the blue motional sidebands of the elastic peak.
Subplots~\ref{fig:5}(b) and~\ref{fig:5}(c) report the details of the
sidebands of the elastic peak. These spectral components correspond to
the photons emitted in the processes where a phonon is created
($\omega_{\rm L} - \nu$) or destroyed ($\omega_{\rm L} + \nu$) in the
mechanical resonator. The motional sideband has a width of the order
of $\propto \Lambda_X^2$, and appears on a broader background with
linewidth $\approx \Gamma$. Our analysis shows that this structure is
due to the fact that mechanical effects are dominated by the strain
coupling $A_X$, which mixes the two excited states. For our parameter
choice, where $\Delta = \nu$, this coupling is weak but resonant so
that the effect of the strain coupling is particularly enhanced.
Figure~\ref{fig:5}(d)-(f) displays corresponding spectra of resonance
fluorescence in presence of the cavity. The significantly
  different features are due to the modified dressed state structure
  because of the strong coupling between cavity and NV center, while
  for both cases the cooling (heating) processes are dominated by
  emission along the transition $|x\rangle\to |g\rangle$
  ($|y\rangle\to |g\rangle$).

The summary of this analysis is that the effect of the optical
resonator on the cooling dynamics can consist in a very small
improvement of the cooling efficiency. This result, which seems to
contrast with previous investigations where the effect of the cavity
on the cooling efficiency was relevant \cite{Zippilli:2005,CEIT:2012},
can be understood when considering that (i) the loss rate of the
resonator and the radiative decay rate of the electronic excitations
have been chosen to be of the same order of magnitude, and (ii) the
cooperativity $C=g^2/\kappa\Gamma\sim 1$, so that the level splitting
induced by the coupling with the resonator is of the order of the loss
rate $\kappa$. Because of (i) the coupling with the resonator gives
rise to an effective level structure where the linewidths of all
excited levels is of the same order of magnitude. Since for sideband
cooling the linewidth determines both the cooling rate as well as the
final temperature, the improvement of the cooling efficiency by
coupling this level structure to a resonator is incremental. Because
of (ii), the level splitting induced by the coupling with the cavity
does not exceed the linewidth of the resonances, so that the regime of
optimal detunings is essentially the same as without the cavity.

\subsection{Dephasing-assisted cooling}

\begin{figure}[!htb]
  \includegraphics[]{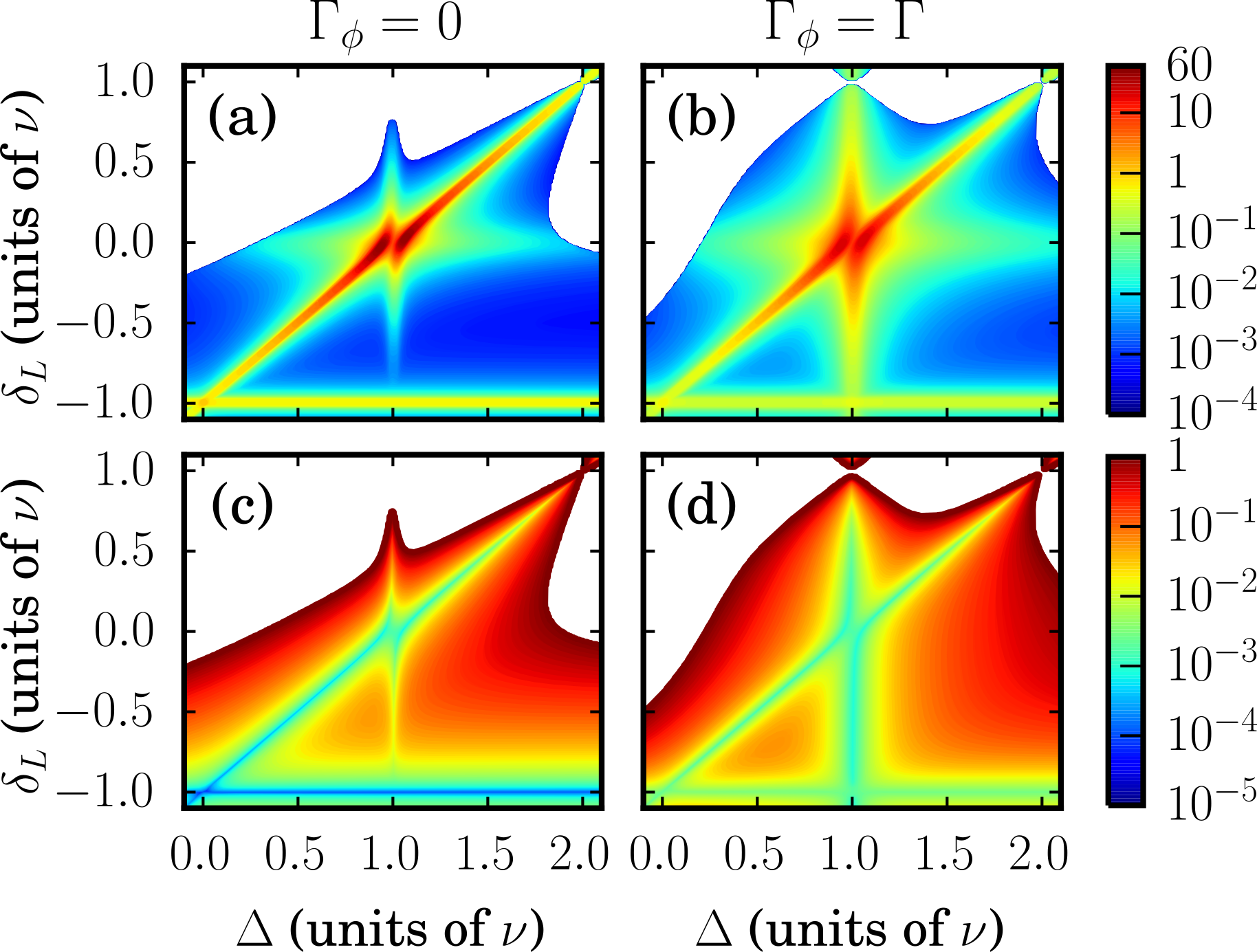}
  \caption{ \label{fig:6} (Color online) Predictions on the cooling
    efficiency extracted from the rate equation, Eq. \eqref{Eq:rate},
    for laser cooling of the mechanical resonator by driving the NV
    center with a laser in absence (left panel) and in presence of
    pure dephasing (right panel).  (a) and (b) show the cooling rate
    $\tilde{\Gamma}$, Eq. \eqref{eq:cooling_rate} in units of
    $\Lambda^2/\nu$, (c) and (d) the asymptotic occupation $N_0$ of
    the vibrational mode, according to Eq. \eqref{eq:mean_phonon}, as
    a function of the excited level splitting $\Delta$ and the laser
    detuning $\delta_{\rm L}$ (in units of $\nu$). The white area are
    heating regions ($\tilde{\Gamma}<0$) or where $N_0>1$. The
    parameters are $\Omega=0.1\nu$, $\Gamma=1.6\times10^{-2}\nu$,
    $\Lambda_I=0,\ \Lambda_X=\Lambda_Z=\chi=\Lambda=0.1\Gamma$, and
    (left panel) $\Gamma_{\phi}=0$, (right panel)
    $\Gamma_{\phi}=\Gamma$.}
\end{figure}

\begin{figure}[!htb]
  \includegraphics[]{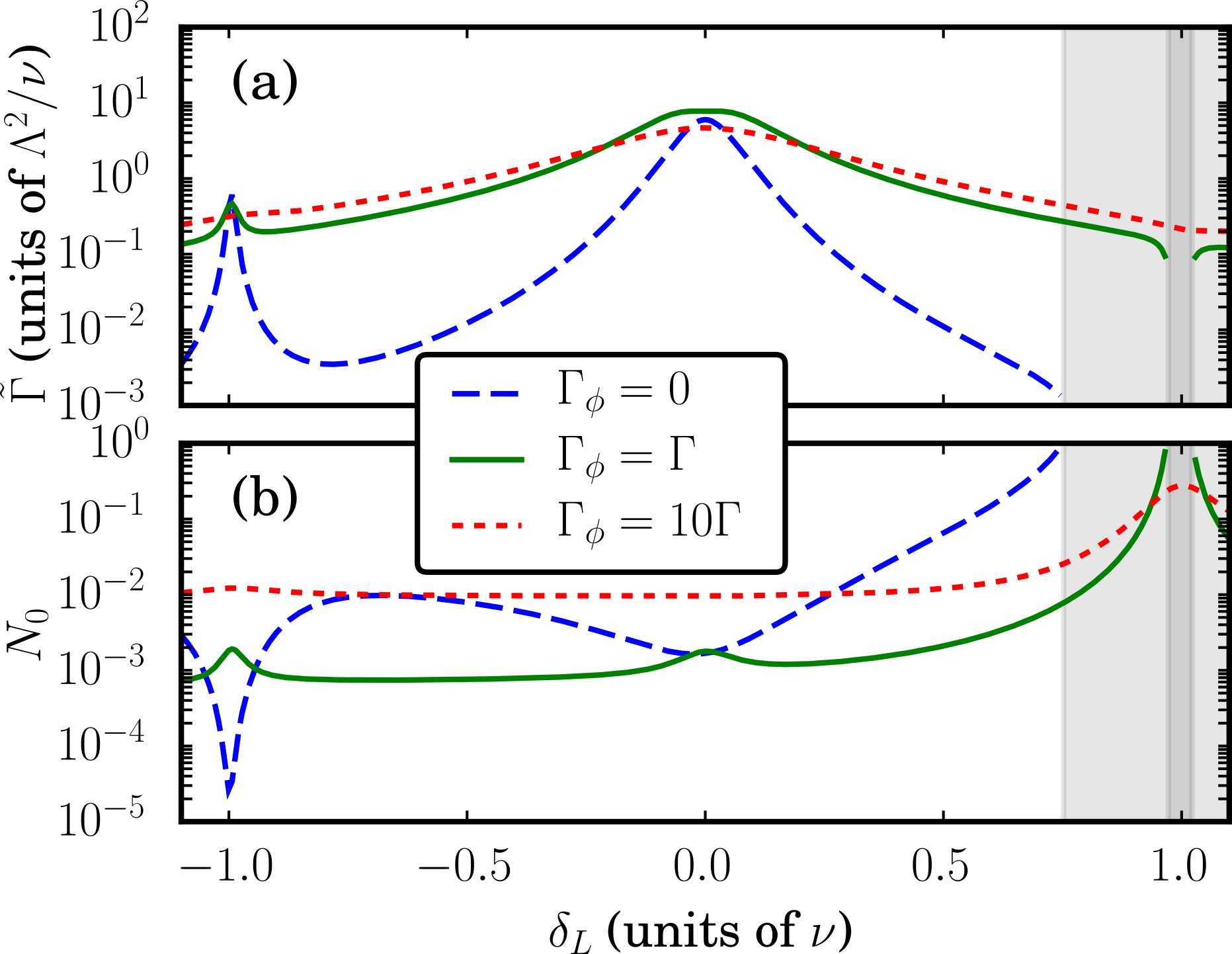}
  \caption{ \label{fig:7} (Color online) (a) Cooling rate
    $\tilde{\Gamma}$ and (b) asymptotic occupation $N_0$ of the
    vibrational mode as a function of $\delta_{\rm L}$ for the same
    parameters as in Fig.~\ref{fig:6} and $\Delta=\nu$. The dashed
    line corresponds to the predictions in absence of dephasing. The
    solid (dotted) line corresponds to the predictions when the
    dephasing rate is $\Gamma_\phi=\Gamma$
    ($\Gamma_\phi=10\Gamma$). The shaded region indicates the regime
    where the resonator is heated by the radiative processes
    ($\tilde{\Gamma}<0$) or where $N_0>1$.}
\end{figure}

We now analyse the effect of other noise sources on the cooling
efficiency, and consider in particular dephasing, which is an
important source of loss of coherence in solid-state systems. We here
discard the coupling with the optical resonator and calculate the
cooling efficiency when $\Gamma_\phi\neq 0$.  Figure~\ref{fig:6}
compares the cooling rate and final occupation for $\Gamma_\phi=0$
(left panel) and $\Gamma_\phi=\Gamma$ (right panel). We observe that
pure dephasing decreases the cooling efficiency when cooling is
achieved by tuning the laser to the red sideband of the dressed
states. Nevertheless, the cooling region is larger and the dependence
on the exact values of the experimental parameters is less
pronounced. Moreover, the cooling performance is enhanced in most
parts of parameter landscape. Figures~\ref{fig:7}(a) and (b) compare
the cuts along the line $\Delta=\nu$: one clearly sees that the case
of $\Gamma_\phi=\Gamma$ outperforms the case when
$\Gamma_\phi=0$. This occurs over almost the full range of
$\delta_{\rm L}$ in terms of both cooling rate and minimal phonon
number. We have checked that the value $\Gamma_\phi=\Gamma$ is close
to the optimal dephasing rate. We also found the range of values in
which the dephasing has a beneficial effect on the cooling spans till
several $\Gamma$ (see dotted line, which shows the predictions for
$\Gamma_\phi=10\Gamma$).

\begin{figure*}[!htb]
  \includegraphics[]{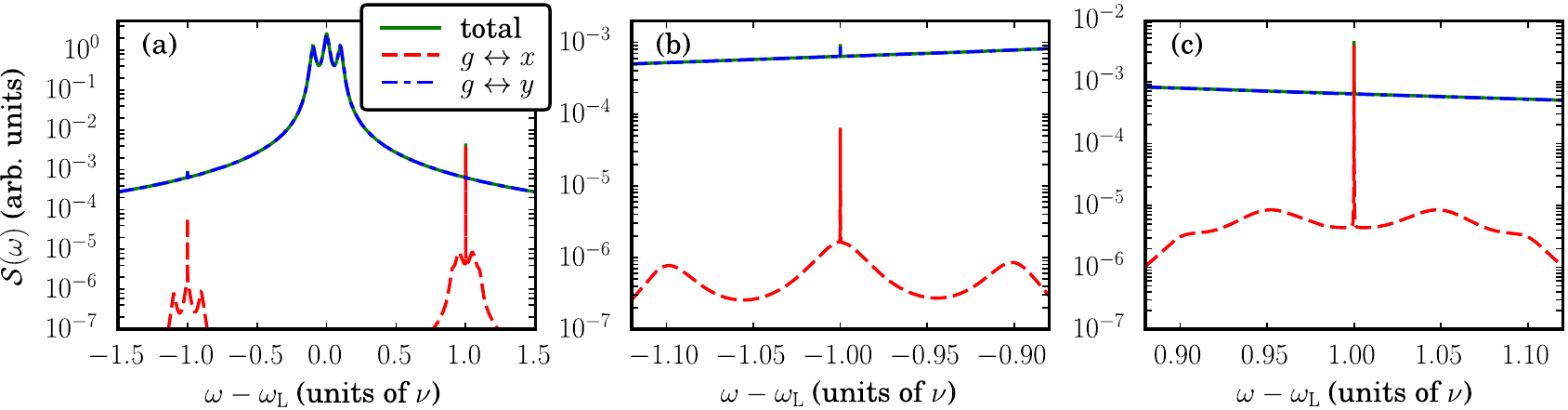}
  \caption{ \label{fig:8} (Color online) Spectrum of light emitted by
    the NV center at the asymptotics of the laser cooling
    dynamics. The parameters are the same as in Fig.~\ref{fig:6}(b)(d)
    (dephasing assisted cooling with $\Gamma_\phi=\Gamma$). The dashed
    (dashed-dotted) line correspond to the emission from the
    transition $g \leftrightarrow x$ ($g \leftrightarrow y$), the
    solid line correspond to the sum of these two contributions. Here,
    we took $\Delta=\nu$ and $\delta_{\rm L}=0$. Panels (b) and (c)
    show the details of the sidebands.}
\end{figure*}

The effect of dephasing is also visible in the spectrum of resonance
fluorescence. We observe in Fig.~\ref{fig:8} for $\delta_{\rm L}=0$ a
broadening of the background at the motional sidebands, which now
scale with $\approx \Gamma + \Gamma_\phi$. This linewidth is indeed
the cooling rate, which results to be enhanced by the presence of pure
dephasing.

We understand this behaviour since pure dephasing increases the width
of the excited states $\ket{x}$ and $\ket{y}$ without increasing their
decay rate. Thus it increases the excitation probability. Since this
cooling scheme is optimal when population is transferred to the
excited state, then pure dephasing leads to larger transition rates,
and thus larger cooling rate. This reasoning works within a certain
parameter interval: dephasing rates exceeding the Rabi frequency, in
fact, tend to suppress population transfer and thus are detrimental.

\begin{figure}[!htb]
  \includegraphics[]{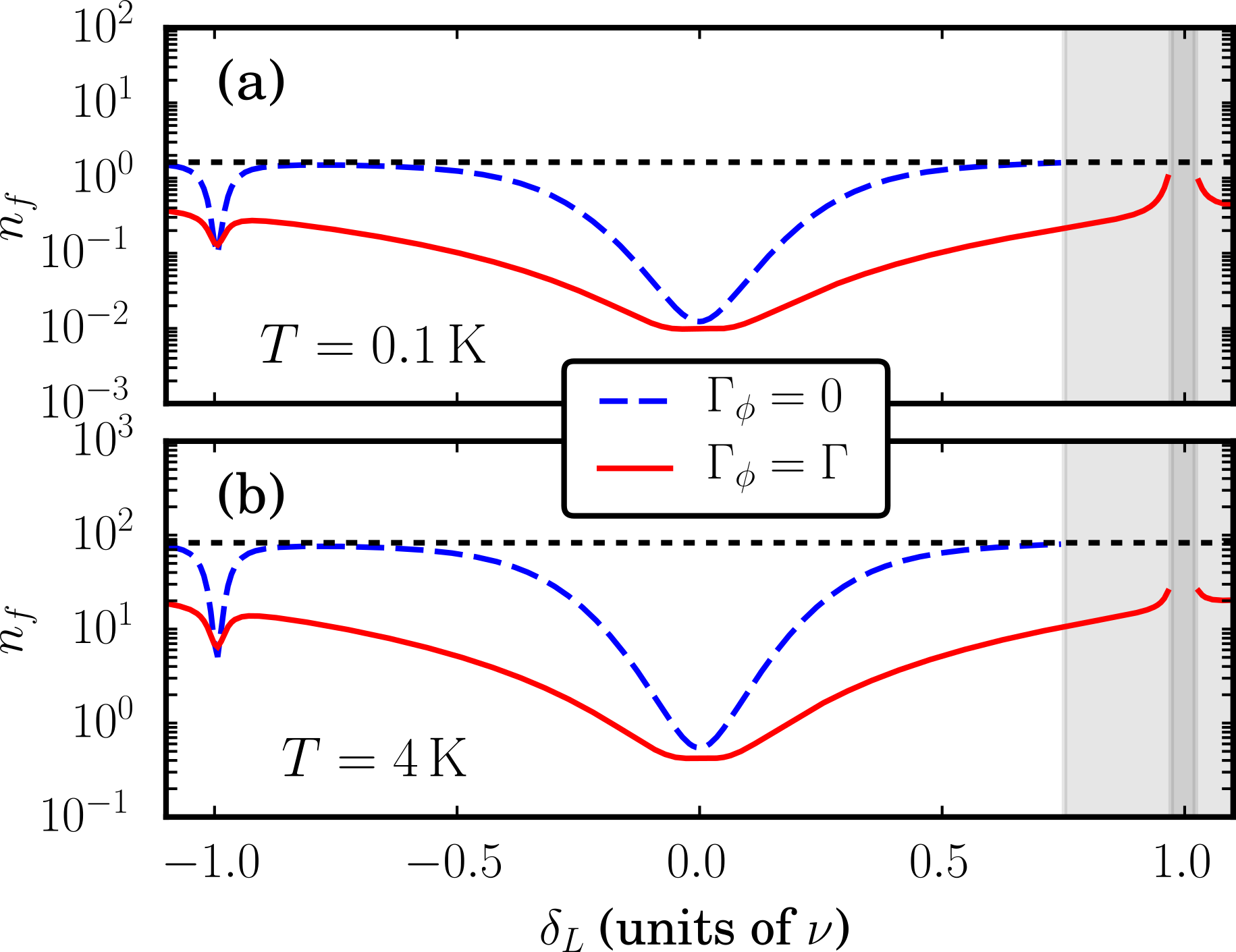}
  \caption{\label{fig:9} (Color online) Final phonon number of the
    mechanical resonator with $\nu=2\pi\times 1\unit{GHz}$ and a
    quality factor $Q=\nu/\gamma=10^7$, Eq.~\eqref{eq:n_f}, for (a)
    $T=0.1\unit{K}$ ($N_{\rm th}\approx1.6$) and (b) $T=4\unit{K}$
    ($N_{\rm th}\approx83$), for the same parameter regime of
    Fig.~\ref{fig:7}(b). The dashed (solid) line corresponds to the
    predictions in absence (presence) of pure dephasing. The black
    dotted lines correspond to $n_f=N_{\rm th}$. The shaded region
    indicates the regime where the resonator is heated by the
    radiative processes.}
\end{figure}

The beneficial role of pure dephasing on the cooling efficiency can be
best illustrated by analysing the final mean occupation for different
temperatures of the bath. Figure~\ref{fig:9} illustrates how dephasing
can improve the cooling efficiency over a large parameter regime,
flattening out the minimum of $n_f$ (Eq.~\eqref{eq:n_f}) as a function
of the frequency of the driving laser.

\section{Discussion and Conclusions}
\label{sec:conclusions}

We have analysed the cooling efficiency of a mechanical resonator
which is laser cooled by the strain coupling with a NV center. The
cooling dynamics is essentially due to the strain coupling with the NV
center and the parameter regime is such that the resolved-sideband
cooling can be performed by driving the NV center electronic
resonances. In this regime we have analysed the effect of the coupling
to an optical resonator, and found that it does only incrementally
improve the cooling efficiency.  We have further shown that pure
dephasing can make the cooling dynamics more robust against parameter
fluctuations, without affecting the overall efficiency, as long as the
dephasing rate does not exceed the driving strength of the laser.

In our analysis the optomechanical coupling was a small effect. It can
be increased in configurations where the cavity is driven: in this
case the optomechanical coupling would cool the resonator according to
the dynamics explored in
Refs.~\cite{WilsonRae2008,Marquardt2007}. Another interesting
possibility is to drive both optical cavity and NV center for large
cooperativity: In this situation phonon excitation or absorption can
be realised by means of two excitation paths, that can interfere. This
interference depends on the relative phase between the lasers and
could be a control parameter for realising multi-wave mixing.

\begin{acknowledgments}
  We thank Peter Rabl, Stefan Sch\"utz, and Christoph Becher for
  useful discussions and helpful comments. This work was partially
  financially supported by the DFG ("Optomechanical cavity Quantum
  Electrodynamics with colour centers in diamond"), the German
  Ministry of Education and Research (BMBF, "Q.com"), the DFG
  Forschergruppe FOR 1493, and the GradUS program of Saarland
  University.
\end{acknowledgments}

\end{document}